\documentclass[preprint,pdf]{iucr}
\usepackage{graphicx}
\usepackage{dcolumn} % align table columns on decimal point
\usepackage{amsmath} 
\usepackage{amssymb}

\usepackage{bm}
\usepackage[separate-uncertainty=true]{siunitx} 

\graphicspath{{figures/}}

\journalcode{J} % Select the journal
\begin{document}
\title{Small-angle neutron scattering by spatially inhomogeneous ferromagnets
with a nonzero average uniaxial anisotropy}
\shorttitle{SANS by macroscopically anisotropic ferromagnets}

\cauthor[a]{V.~D.}{Zaporozhets}{vladdz@donfti.ru}{address if different from \aff}
\author[b]{Y.}{Oba}
\author[c]{A.}{Michels}
\author[a,d]{K.~L.}{Metlov}

\aff[a,d]{Donetsk Institute for Physics and Technology, Donetsk 83114, \country{Ukraine}}
\aff[b]{Materials Sciences Research Center, Japan Atomic Energy Agency, 2-4 Shirakata, Tokai, Ibaraki, 319-1195, \country{Japan}}
\aff[c]{Department of Physics and Materials Science, University of Luxembourg, 162A Avenue de la Faiencerie, L-1511 Luxembourg, \country{Grand Duchy of Luxembourg}}
\aff[d]{Institute for Numerical Mathematics RAS, 8 Gubkina str., 119991 Moscow GSP-1, \country{Russia}}

\shortauthor{Zaporozhets, Oba, Michels, and Metlov}

%\vita{Author's biography}

\maketitle

% useful macros
\let\vec=\bm

\def\ud{\,{\mathrm{d}}}
\def\uiint{\int\!\!\!\int}
\def\uiiint{\int\!\!\!\int\!\!\!\int}
\def\uRe{{\mathrm{Re}}\,}
\def\uIm{{\mathrm{Im}}\,}

% symbols
\def\uAx{\widetilde{A}_\mathrm{X}}
\def\uAy{\widetilde{A}_\mathrm{Y}}
\def\uAxp{\widetilde{A}_\mathrm{X}^\prime}
\def\uAyp{\widetilde{A}_\mathrm{Y}^\prime}

% scattering length per Borh magneton
\def\uA{A} % exchange stiffness C=2A
\def\ubH{b_\mathrm{H}}
% magnetic induction vector
\def\uvB{\vec{B}} 
\def\uC{C} % exchange stiffness C=2A
\def\uCONV{\oplus}
\def\uvd{\vec{d}}     % local anisotropy axis
\def\uvdp{\vec{d}^\prime}
\def\uFdxp{\widetilde{d}_\mathrm{X}^\prime}
\def\uFdyp{\widetilde{d}_\mathrm{Y}^\prime}
\def\uFdzp{\widetilde{d}_\mathrm{Z}^\prime}
\def\ue{e}
\def\ueX{\uO\uX}
\def\ueY{\uO\uY}
\def\ueZ{\uO\uZ}
\def\ueXp{\uO\uXp}
\def\ueYp{\uO\uYp}
\def\ueZp{\uO\uZp}
\def\uE{E}
\def\ugammaB{\gamma_\mathrm{B}}
\def\uGrad{\vec{\nabla}}
\def\uh{h}            % normalized magnetic field
\def\uvh{\vec{h}}
\def\uvhp{\vec{h}^\prime}
\def\uhq{h_q}         % the external field enhanced by the exchange
\def\uhqp{h_q^\prime}
\def\uhqpsq{h_q^{\prime2}}
% the normalized external field components
% in primed coordinate system
\def\uhxp{h_\mathrm{X}^\prime}
\def\uhyp{h_\mathrm{Y}^\prime}
\def\uhzp{h_\mathrm{Z}^\prime}
\def\uH{H}            % external magnetic field
\def\uvH{\vec{H}}
\def\uvHp{\vec{H}^\prime}
\def\uvHD{\vec{H}_\mathrm{D}} % demag. field
\def\uHK{H_\mathrm{K}}  % anisotropy field
\def\uvHeff{\vec{H}_\text{eff}} % effective field
\def\uvHeffp{\vec{H}_\text{eff}^\prime}
\def\uvHeffXp{\vec{H}_\text{X, eff}^\prime}
\def\uvHeffYp{\vec{H}_\text{Y, eff}^\prime}
\def\uvHeffZp{\vec{H}_\text{Z, eff}^\prime}
\def\uvHeffA{\vec{H}_\text{eff}^\mathrm{A}}
\def\uvFHeff{\widetilde{\vec{H}}_\text{eff}}
\def\uvFHeffp{\widetilde{\vec{H}}_\text{eff}^\prime}
\def\uvFHeffXp{\widetilde{H}_\text{eff, X}^\prime}
\def\uvFHeffYp{\widetilde{H}_\text{eff, Y}^\prime}
\def\uvFHeffZp{\widetilde{H}_\text{eff, Z}^\prime}
\def\uHeffi{H^\text{eff}_i} % component of the effective field i=X,Y,Z
\def\uIr{I}
\def\uImr{I_\mathrm{M}}
\def\uIkr{I_\mathrm{K}}
\def\uFIr{\widetilde{I}}
\def\uFImr{\widetilde{I}_\mathrm{M}}
\def\uFIkr{\widetilde{I}_\mathrm{K}}
\def\uK{K}
\def\uKZ{K_0}
\def\uL{L}
\def\uLe{L_\mathrm{E}}
\def\uLZ{L_0} % the average exchange length
\def\uvM{\vec{M}}     % magnetization vector
\def\uvMp{\vec{M}^{\prime}} % m. v. primed
\def\uM{M}
\def\uMi{M_i}  % i=X,Y,Z
\def\uMx{M_\mathrm{X}}
\def\uMy{M_\mathrm{Y}}
\def\uMz{M_\mathrm{Z}}
\def\uMxp{M_\mathrm{X}^{\prime}}
\def\uMyp{M_\mathrm{Y}^{\prime}}
\def\uMzp{M_\mathrm{Z}^{\prime}}
\def\uMxpI{M_\mathrm{X}^{\prime(1)}}
\def\uMypI{M_\mathrm{Y}^{\prime(1)}}
\def\uMzpI{M_\mathrm{Z}^{\prime(1)}}
\def\uFM{\widetilde{M}}
\def\uvFM{\widetilde{\vec{M}}}
\def\uvFMp{\widetilde{\vec{M}}^{\prime}}
\def\uFMx{\widetilde{M}_\mathrm{X}}
\def\uFMy{\widetilde{M}_\mathrm{Y}}
\def\uFMz{\widetilde{M}_\mathrm{Z}}
\def\uFMxp{\widetilde{M}_\mathrm{X}^{\prime}}
\def\uFMyp{\widetilde{M}_\mathrm{Y}^{\prime}}
\def\uFMzp{\widetilde{M}_\mathrm{Z}^{\prime}}
% normalized magnetization
\def\uvm{\vec{m}}     % magnetization vector
\def\uvmp{\vec{m}^{\prime}} % m. v. primed
\def\umi{m_i}  % i=X,Y,Z
\def\umx{m_\mathrm{X}}
\def\umy{m_\mathrm{Y}}
\def\umz{m_\mathrm{Z}}
\def\umxp{m_\mathrm{X}^{\prime}}
\def\umyp{m_\mathrm{Y}^{\prime}}
\def\umzp{m_\mathrm{Z}^{\prime}}
\def\umxpI{m_\mathrm{X}^{\prime(1)}}
\def\umypI{m_\mathrm{Y}^{\prime(1)}}
\def\umzpI{m_\mathrm{Z}^{\prime(1)}}
\def\uFm{\widetilde{m}}
\def\uvFm{\widetilde{\vec{m}}}
\def\uvFmp{\widetilde{\vec{m}}^{\prime}}
\def\uFmx{\widetilde{m}_\mathrm{X}}
\def\uFmy{\widetilde{m}_\mathrm{Y}}
\def\uFmz{\widetilde{m}_\mathrm{Z}}
\def\uFmxp{\widetilde{m}_\mathrm{X}^{\prime}}
\def\uFmyp{\widetilde{m}_\mathrm{Y}^{\prime}}
\def\uFmzp{\widetilde{m}_\mathrm{Z}^{\prime}}
\def\uFmxpI{\widetilde{m}_\mathrm{X}^{\prime(1)}}
\def\uFmypI{\widetilde{m}_\mathrm{Y}^{\prime(1)}}
\def\uFmzpI{\widetilde{m}_\mathrm{Z}^{\prime(1)}}
\def\uN{N}
\def\uFN{\widetilde{N}}
\def\uMs{M_\mathrm{S}}% saturation magnetization
\def\uMZ{M_0}% average saturation magnetization
\def\uFMs{\widetilde{M}_\mathrm{S}}
\def\umuZ{\mu_0}
\def\uO{O}       % Cartesian coordinates origin
\def\uRotation{\hat{R}}
\def\uQ{Q}            % global anis. Q-factor
\def\uQVP{Q_\text{VP}}       % Q for Vitroperm
\def\uQNi{Q_\text{HPT--Ni}}  % Q for HPT-Ni
\def\uq{q}            % q-vector magnitude
% response functions
\def\uRH{R_\mathrm{H}}
\def\uRM{R_\mathrm{M}}
\def\uRHPerp{R_\mathrm{H}^{\perp}}
\def\uRHPar{R_\mathrm{H}^{\parallel}}
\def\uRMPerp{R_\mathrm{M}^{\perp}}
\def\uRMPar{R_\mathrm{M}^{\parallel}}
\def\uRHXPerp{R_\mathrm{H}^{\perp,\mathrm{X}}}
\def\uRMXPerp{R_\mathrm{M}^{\perp,\mathrm{X}}}
\def\uRHYPerp{R_\mathrm{H}^{\perp,\mathrm{Y}}}
\def\uRMYPerp{R_\mathrm{M}^{\perp,\mathrm{Y}}}
\def\uRHXYPerp{R_\mathrm{H}^{\perp,\mathrm{X/Y}}}
\def\uRMXYPerp{R_\mathrm{M}^{\perp,\mathrm{X/Y}}}
\def\uRHXPar{R_\mathrm{H}^{\parallel,\mathrm{X}}}
\def\uRMXPar{R_\mathrm{M}^{\parallel,\mathrm{X}}}
\def\uRHYPar{R_\mathrm{H}^{\parallel,\mathrm{Y}}}
\def\uRMYPar{R_\mathrm{M}^{\parallel,\mathrm{Y}}}
\def\uRHXYPar{R_\mathrm{H}^{\parallel,\mathrm{X/Y}}}
\def\uRMXYPar{R_\mathrm{M}^{\parallel,\mathrm{X/Y}}}
\def\uvq{\vec{q}}     % q-vector
\def\uvqPar{\vec{q}^\parallel}
\def\uvqPerp{\vec{q}^\perp}
\def\uvqp{\vec{q}^{\prime}}  % q-vector primed
\def\uqx{q_\mathrm{X}}
\def\uqy{q_\mathrm{Y}}
\def\uqz{q_\mathrm{Z}}
\def\uqxp{q_\mathrm{X}^{\prime}}
\def\uqyp{q_\mathrm{Y}^{\prime}}
\def\uqzp{q_\mathrm{Z}^{\prime}}
\def\uphi{\varphi}    % asimuthal angle
\def\uvr{\vec{r}}     % radius-vector
\def\uvs{\vec{s}}     % global anisotropy axis
\def\uvsp{\vec{s}^\prime} % glb. an. axis primed
% components of the global anisotropy
% director in primed coordinate system
\def\uspx{s_\mathrm{X}^{\prime}}
\def\uspy{s_\mathrm{Y}^{\prime}}
\def\uspz{s_\mathrm{Z}^{\prime}}
% squared components of the global anisotropy
% director in primed coordinate system
\def\uspxsq{s_\mathrm{X}^{\prime 2}}
\def\uspysq{s_\mathrm{Y}^{\prime 2}}
\def\uspzsq{s_\mathrm{Z}^{\prime 2}}
\def\uSH{S_\mathrm{H}}
\def\uSM{S_\mathrm{M}}
\def\uSigmadInline{\mathrm{d} \Sigma/\mathrm{d}\Omega}
\def\uSigmadRES{\frac{\mathrm{d} \Sigma_\mathrm{res}}{\mathrm{d}\Omega}}
\def\uSigmadRESInline{\mathrm{d} \Sigma_\mathrm{res}/\mathrm{d}\Omega}
\def\uSigmadSMInline{\mathrm{d} \Sigma_\mathrm{M}/\mathrm{d}\Omega}
\def\uSigmadPar{\frac{\mathrm{d} \Sigma^\parallel}{\mathrm{d} \Omega}}  % total parallel differential cross-section
\def\uSigmadPerp{\frac{\mathrm{d} \Sigma^\perp}{\mathrm{d} \Omega}}  % total perpendicular differential cross-section
% spin misalignment parallel differential cross-section
\def\uSigmadPerpInline{\mathrm{d} \Sigma^\perp/\mathrm{d} \Omega}
\def\uSigmadSMPar{\frac{\mathrm{d} \Sigma_\mathrm{M}^\parallel}{\mathrm{d} \Omega}}  
\def\uSigmadSMParInline{\mathrm{d} \Sigma_\mathrm{M}^\parallel/\mathrm{d} \Omega}  
% spin misalignment perpendicular differential cross-section
\def\uSigmadSMPerp{\frac{\mathrm{d} \Sigma_\mathrm{M}^\perp}{\mathrm{d}\Omega}} 
\def\uSigmadSMPerpInline{\mathrm{d} \Sigma_\mathrm{M}^\perp/\mathrm{d}\Omega} 
\def\uSigmadSMPerpp{\frac{\mathrm{d} \Sigma_\mathrm{M}^{\perp\prime}}{\mathrm{d} \Omega}}
\def\uSigmadSMPerppInline{\mathrm{d}\Sigma_\mathrm{M}^{\perp\prime}/\mathrm{d}\Omega}
\def\uSigmadSMPerpY{\frac{\mathrm{d} \Sigma_\mathrm{M}^{\perp,\mathrm{Y}}}{\mathrm{d}\Omega}}
\def\uSigmadSMPerpZ{\frac{\mathrm{d} \Sigma_\mathrm{M}^{\perp,\mathrm{Z}}}{\mathrm{d}\Omega}}

\def\utheta{\theta}   % polar angle
\def\uV{V}            % scattering volume
\def\uX{X}            % Cartesian X coordinate
\def\uY{Y}            % Cartesian Y coordinate
\def\uZ{Z}            % Cartesian Z coordinate
\def\uXp{X^\prime}    % Cartesian X' coordinate
\def\uYp{Y^\prime}    % Cartesian Y' coordinate
\def\uZp{Z^\prime}    % Cartesian Z' coordinate
% components of the q-vector director
\def\uxq{x_{\vec{q}}}
\def\uyq{y_{\vec{q}}}
\def\uzq{z_{\vec{q}}}
\def\uxqsq{x_{\vec{q}}^2}
\def\uyqsq{y_{\vec{q}}^2}
\def\uzqsq{z_{\vec{q}}^2}
% the same, but in primed coordinate system
\def\uxqp{x_{\vec{q}^\prime}}
\def\uyqp{y_{\vec{q}^\prime}}
\def\uzqp{z_{\vec{q}^\prime}}
\def\uxqpsq{x_{\vec{q}^\prime}^2}
\def\uyqpsq{y_{\vec{q}^\prime}^2}
\def\uzqpsq{z_{\vec{q}^\prime}^2}

\begin{abstract}
Micromagnetic small-angle neutron scattering theory is well established for analyzing spin-misalignment scattering data of bulk ferromagnets. Here, this theory is extended to allow for a global uniaxial magnetic anisotropy (texture) of the material, in addition to the already included random zero-average local anisotropy. Macroscopic cross-sections and spin-misalignment response functions are computed analytically for several practically relevant mutual anisotropy and external magnetic field orientations in both parallel and perpendicular scattering geometries for field magnitudes both above and below the rotational saturation. Some of these expressions are tested on published experimental data of magnetic-field-annealed Vitroperm and plastically-deformed Ni, allowing to determine the corresponding global uniaxial anisotropy quality factors.
\end{abstract}

\section{INTRODUCTION}

Small-angle neutron scattering (SANS) is one of the most powerful techniques for studying magnetic textures of bulk magnets on the scale between a few and a few hundred of nanometers. To interpret the scattering cross-section data it is useful to combine the theory of neutron scattering (describing the passage of neutrons through a magnetized sample) and the theory of micromagnetics (describing the formation of the magnetization texture in the material). Such a combination in the form of the micromagnetic SANS theory claimed a number of successes for the description of SANS experiments and for the extraction of useful information regarding the samples magnetic microstructure~\cite{michelsSANSbook2021}.

Current micromagnetic SANS theory, however, deals only with macroscopically isotropic magnets. Such systems can have small local fluctuating magnetic anisotropies, but their magnitude and direction average out, so that the sample as a whole remains statistically isotropic. There is abundance of such samples, but there are also plenty of material systems which do have a non-zero global anisotropy (texture). Moreover, the anisotropy can be induced in an originally statistically isotropic magnet artificially, \textit{e.g.}\ by annealing it in the presence of a magnetic field, applying mechanical stress, or subjecting it to severe plastic deformation.

The purpose of this paper is to extend the micromagnetic SANS theory to systems with a global uniaxial anisotropy. Starting from the relevant micromagnetic Hamiltonian, we compute the spin-misalignment SANS cross-sections analytically for many practically interesting cases of mutual orientations between the external magnetic field and the anisotropy axis. The results reveal new effects and quantitatively describe well-known observations. We test the theory by applying it to previously published experimental SANS data.

The work is organized as follows:~In Section~\ref{sec:SANS} we introduce the nomenclature and the expressions for the SANS cross-sections in terms of the Fourier image of the magnetization distribution of the sample. Section~\ref{sec:MicroMag} contains the solution of the micromagnetic problem for the magnetization distribution of a sample with global anisotropy. When the magnetic field is perpendicular to the global anisotropy axis, there are two distinct cases:~the {\em small-field} limit (when the direction of the average magnetization is determined by the balance of the external field and the anisotropy torques) and a simpler {\em high-field} limit (when the average magnetization is aligned strictly along the external field direction, but its small fluctuations are still influenced by the global anisotropy of the sample). After performing the averaging of the SANS cross-sections over the orientations and realizations of the random magnetic parameter fluctuations in the scattering volume, the resulting macroscopic cross-section expressions are presented in Sections~\ref{sec:lowfield} and \ref{sec:highfield} in the low and high-field limits for several chosen mutual orientations between the anisotropy axis and the external magnetic field. In Section~\ref{sec:expdata} we apply our theory to the interpretation of unpolarized SANS data on a field-annealed nanocrystalline bulk Vitroperm metallic glass and on a pure Ni sample, subjected to severe plastic deformation. Finally, Section~\ref{sec:summary} summarizes our main results.

\section{MAGNETIC SMALL-ANGLE NEUTRON SCATTERING}\label{sec:SANS}

\begin{figure}
\label{fig:SANSscheme}
\includegraphics[width=1.0\textwidth]{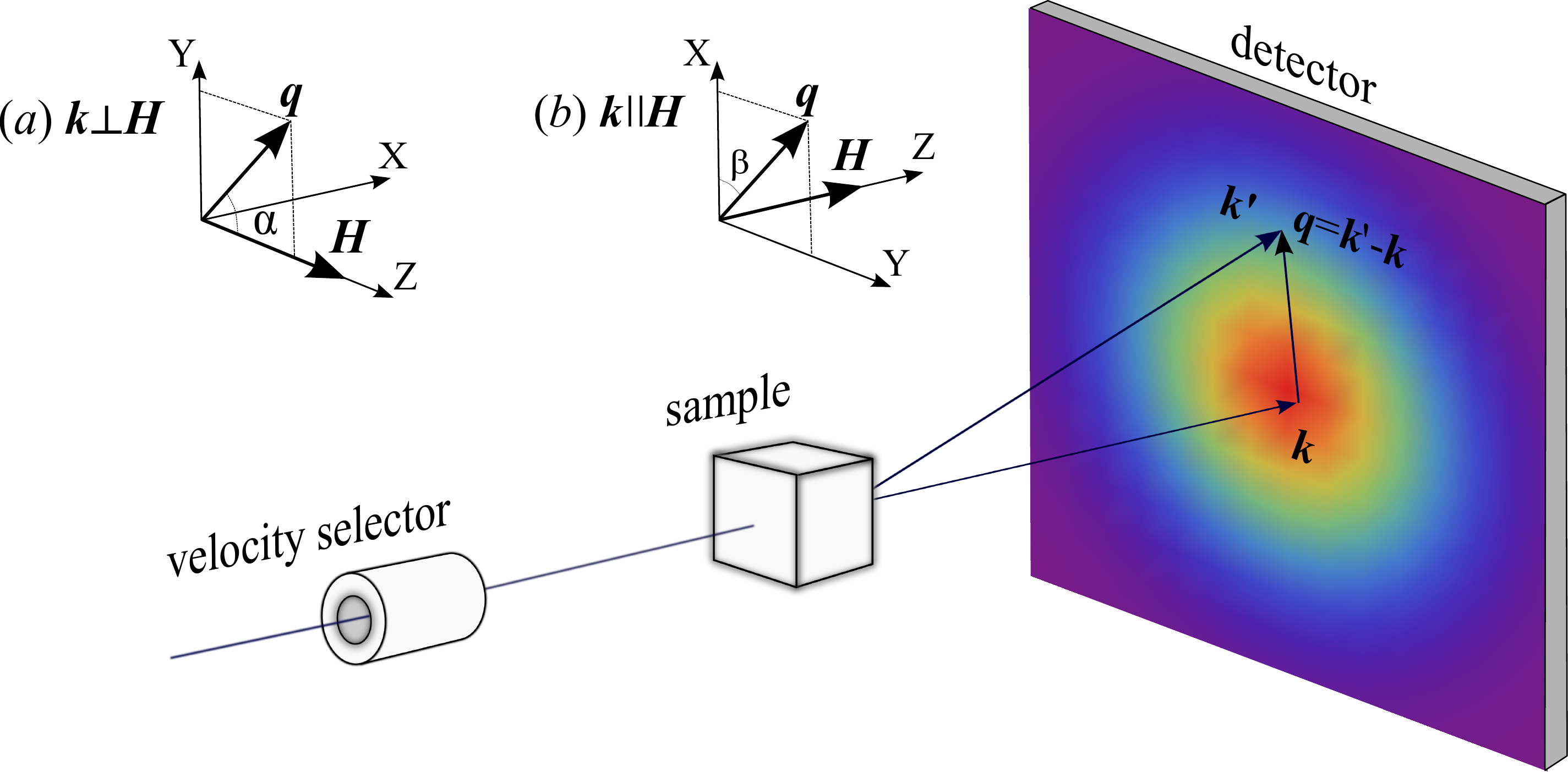}
\caption{Typical setup of a magnetic SANS experiment and the definition of the scattering vector $\uvq$. The insets depict the coordinate systems used for the perpendicular (\textit{a}) and the parallel (\textit{b}) scattering geometries.}
\end{figure}

A typical scheme of a small-angle neutron scattering (SANS) experiment is shown in Fig.~\ref{fig:SANSscheme}. Each of the incident neutrons with a well-defined energy and wave vector $\vec{k}$ is scattered by the sample and takes on a new wave vector $\vec{k}'$. The difference $\uvq = \vec{k}-\vec{k}'$ is called the scattering or momentum-transfer vector. Magnetic SANS experiments are usually performed in the presence of an applied magnetic field $\uvH$. Out of all the possible mutual arrangements of the vectors $\vec{k}$ and $\uvH$, two scattering geometries are usually employed:~the perpendicular one ($\vec{k}\perp\uvH$) and the parallel one ($\vec{k}\parallel\uvH$). We can associate a Cartesian coordinate system with each of the geometries in such a way that their $\ueZ$~axis coincides with the direction of the applied magnetic field $\uvH$ and either the $\ueX$ [in the perpendicular geometry, Fig.~\ref{fig:SANSscheme}(\textit{a})] or the $\ueZ$ [in the parallel geometry, Fig.~\ref{fig:SANSscheme}(\textit{b})] axis coincides with the wave vector of incident neutrons. In SANS, the vector $\uvq$ is assumed to lie in the detector plane and is usually parametrized as $\uvq^{\perp} = \uq \{0, \sin \alpha, \cos \alpha\}$ or $\uvq^{\parallel} = \uq\{\cos \beta, \sin \beta, 0\}$ in the perpendicular and the parallel geometry, respectively.

Assume that the sample is a magnetically-ordered substance (ferro- or ferrimagnet) at a sufficiently low temperature, far from the order-disorder phase transition (and the compensation temperature, if it is a ferrimagnet). Such a medium can be characterized by a nonzero local magnetization vector $\uvM$, with $|\uvM|=\uMs$ being the saturation magnetization of the material. At the mesoscopic level, the spatial distribution of the local magnetization (magnetic texture) can be represented by a continuous vector field $\uvM(\uvr)$, where $\uvr$ is a spatial coordinate.

If the material is infinite, isotropic, and uniform, its equilibrium magnetization will be oriented along the direction of the external field $\uvH$. Otherwise, there will be some spatial variation of the magnetization, which will manifest itself, in particular, via the magnetic scattering of neutrons at nonzero scattering vectors $\uvq$. The corresponding  total (magnetic and nuclear) macroscopic scattering cross-section $\uSigmadInline$ can be expressed as the sum $\uSigmadInline=\uSigmadRESInline+\uSigmadSMInline$, where the first term (residual scattering) is magnetic-field-independent and the second term (spin-misalignment scattering) vanishes under the condition of magnetic saturation (very large external magnetic field magnitude). In the perpendicular (superscript $\perp$) and parallel ($\parallel$) scattering geometries the unpolarized spin-misalignment scattering cross-sections can be expressed as follows~\cite{michelsSANSbook2021}:
\begin{eqnarray}
\uSigmadSMPerp =\uV \ubH^2 \left[|\uFMx|^2+|\uFMy|^2\cos^2\alpha+ (|\uFMz|^2\!-\!|\uFMs|^2) \sin^2\alpha - \uRe(\overline{\uFMy} \uFMz)\sin2\alpha \right],
\label{eq:perpcrossect} \\
\uSigmadSMPar = \uV \ubH^2 \left[|\uFMx|^2 \sin^2 \beta +|\uFMy|^2 \cos^2\beta + (|\uFMz|^2\!-\!|\uFMs|^2) - \uRe(\overline{\uFMy} \uFMx)\sin2\beta \right],
\label{eq:parcrossect}
\end{eqnarray}
where $\ubH=\SI{2.906e8}{\per\ampere\per\meter}$ is the magnetic scattering length per Bohr magneton, $\uV$ denotes the scattering volume, the polar angles $\alpha$ and $\beta$ of the scattering vector $\uvq$ in the detector plane are schematically depicted in the insets of Fig.~\ref{fig:SANSscheme}, $\uvM=\{\uMx,\uMy,\uMz\}$ is the magnetization vector inside the material with Cartesian components $M_{\mathrm{X,Y,Z}}$ (in the coordinate system for a particular scattering geometry), lines above the symbols stand for the complex conjugate, and tildes denote the discrete Fourier transform:
\begin{equation}
\widetilde{F}(\uvq) = \frac{1}{\uV} \uiiint_\uV F(\uvr) e^{-\imath \uvq\cdot\uvr} \ud^3 \uvr,
\qquad
F(\uvr) = \sum_{\uvq} \widetilde{F}(\uvq) e^{\imath \uvq\cdot\uvr} .
\label{eq:Fourier}
\end{equation}
The integral in~\eqref{eq:Fourier} is taken over a representative cube of the material $\uV = \uL\times \uL\times \uL$, corresponding to the coherence volume of the neutron beam, which is considered to be periodically repeating. The Cartesian components of $\uvq=\{\uqx,\uqy,\uqz\}$ therefore take on all values which are integer multiples of $2\pi/\uL$. Because of this different definition of the Fourier transform, the expressions~\eqref{eq:perpcrossect} and~\eqref{eq:parcrossect} differ by the factor $\uV^2/(8\pi^3)$ from the ones in the original work~\cite{michels2013}.

Thus, computing the SANS cross-sections~\eqref{eq:perpcrossect} and~\eqref{eq:parcrossect} boils down to finding the Fourier images of the magnetization vector components $\{\uFMx,\uFMy,\uFMz\}$. The latter can be obtained by solving the corresponding micromagnetic problem.

\section{MICROMAGNETICS OF A WEAKLY INHOMOGENEOUS YET GLOBALLY ANISOTROPIC MAGNET}\label{sec:MicroMag}

Micromagnetics~\cite{Brown_micromagnetics_1963} is based on the minimization of the total energy of the magnet $\uE$, whose local magnetic moments are subject to various well-known magnetic interactions. Specifically, in this work, similarly to~\citeasnoun{MM2015}, we assume that the total energy can be expressed as an integral over the energy density $\uE=\uiiint_\uV \ue \ud^3\uvr$ with
\begin{eqnarray}
	\ue & = & \frac{\uC(\uvr)}{2}\sum_{i=\uX,\uY,\uZ} \left[\uGrad\frac{\uMi(\uvr)}{\uMs(\uvr)}\right]^2 -\umuZ \uvM(\uvr) \cdot \uvH -\frac{1}{2}\umuZ\uvM(\uvr) \cdot \uvHD (\uvr,\{\uvM(\uvr)\})-\nonumber\\
	& &  - \uK(\uvr)\left[\frac{\uvM(\uvr)}{\uMs(\uvr)} \cdot \uvd(\uvr)\right]^2 - \uKZ \left[\frac{\uvM(\uvr)}{\uMs(\uvr)} \cdot \uvs\right]^2,\label{eq:energy}
\end{eqnarray}
where $\uC(\uvr)$ is the position-dependent exchange stiffness, $\uGrad = \{\partial/\partial \uX, \partial/\partial \uY, \partial/\partial \uZ\}$, $\uvH=\{0,0,\uH\}$ is the external magnetic field (always aligned along the $\ueZ$~axis as per previously discussed convention), $\uvHD$ is the demagnetizing field, created by the magnetization distribution $\uvM(\uvr)$, which explicitly depends on $\uvr$ and has a functional dependence on the whole $\uvM(\uvr)$ vector field via Maxwell equations, $\uKZ$ and $\uK(\uvr)$ denote the constant global and small spatially fluctuating local anisotropy parameters, $\uvs$ and $\uvd(\uvr)$ are the corresponding anisotropy axis directors ($|\uvd|=|\uvs|=1$), and $\umuZ$ is the permeability of vacuum. The anisotropy is of the easy-axis or easy-plane type, depending on whether $\uKZ > 0$ or $\uKZ < 0$. We also assume that the saturation magnetization is weakly fluctuating:
\begin{equation}
	\uMs(\uvr) = \uMZ[1+\uImr(\uvr)], 
\end{equation} 
where the magnitude of $\uImr(\uvr)\ll1$ is a small quantity with a zero spatial average $\langle \uImr(\uvr)\rangle = 0$, so that $\uMZ = \langle \uMs(\uvr)\rangle$ is the average saturation magnetization of the magnet. The quality factor of the fluctuating anisotropy $\uIkr(\uvr) = 2 \uK(\uvr)/[\umuZ \gamma_B \uMs^2(\uvr)] \ll1$ is assumed to be small and of the same order as $\uImr$ with its average being zero: $\langle\uIkr(\uvr)\rangle=0$. To cover several systems of magnetic units, the magnetic induction is defined here as $\uvB = \umuZ(\uvH + \ugammaB \uvM)$, where $\ugammaB=1$ in SI units, and $\ugammaB=4\pi$, $\umuZ=1$ in CGS units~\cite{Aharoni96}.

The model~\eqref{eq:energy} with $\uKZ=0$ was already studied by~\citeasnoun{michels2013}, \citeasnoun{MM2015} and in many other papers under a wide spectrum of assumptions about the random material parameters fluctuations (random defects). Global uniaxial anisotropy $\uKZ\neq0$, which is new in~\eqref{eq:energy} in the context of magnetic SANS, can have several physical origins. It may appear due to heat treatment in a magnetic field (magnetic annealing). It can be the result of an applied uniaxial mechanical stress or annealing of a material that is simultaneously subjected to an applied stress (stress annealing). Or, it can be imprinted into the material by severe plastic deformation (\textit{e.g.}\ by high-pressure-torsion).

The equilibrium magnetization vector distribution, minimizing~\eqref{eq:energy} under the constraint $|\uvM|=\uMs$, is a solution to the Brown's equations~\cite{Brown_micromagnetics_1963,Aharoni96}:
\begin{equation} \label{eq:Brown}
	\uvHeff(\uvr) \times \uvM(\uvr) = 0,
\end{equation}
where the cross denotes the vector product. The effective magnetic field $\uvHeff(\uvr) $ is defined as the functional derivative of the ferromagnet's energy-density functional $\uE(\{\uvM(\uvr)\})$ over the magnetization vector field
\begin{equation}
        \label{eq:HeffDEF}
	\uHeffi(\uvr) = -\frac{1}{\umuZ} \frac{\delta \ue}{\delta \uMi} = -\frac{1}{\umuZ}\left(\frac{\partial \ue}{\partial \uMi}- \uGrad \cdot \frac{\partial \ue}{\partial \uGrad \uMi}\right).
\end{equation} 
Linearity of the variational derivative allows to compute the contribution of each term in~\eqref{eq:energy} separately. They were already published earlier~\cite{MM2015}. Let us give here only the expression for the effective field associated with the magnetic anisotropy [the last two terms in~\eqref{eq:energy}]:
\begin{equation}
        \label{eq:HeffA}
	\uvHeffA = \gamma_B \uIkr(\uvr) [\uvM(\uvr)\cdot \uvd(\uvr)] \uvd(\uvr) + \ugammaB \uQ (1-2 \uImr) [\uvM(\uvr) \cdot \uvs] \uvs,
\end{equation} 
where
\begin{equation} \label{eq:Q}
       \uQ = \frac{2 \uKZ}{\umuZ \ugammaB \uMZ^2}
\end{equation}
is the global anisotropy quality factor. Note that the expression~\eqref{eq:HeffA} is only valid up to the first order in small quantities $\uIkr$ and $\uImr$, higher-order terms of this expansion are ignored. The contribution $\propto2\uImr$ in the last term appears due to the series expansion of $\uMs^2(\uvr)$ in the denominator of the corresponding term in~\eqref{eq:energy}. In this work, we will limit ourselves only to the second-order theory (first-order solution of the micromagnetic problem, second-order contribution to the magnetic SANS cross-sections) in the amplitude of small material parameters fluctuations. Consequently, the exchange length $\uLe$ can also be considered as a nonfluctuating constant $\uLe^2=\uLZ^2=\langle\uC(\uvr)/(\umuZ \ugammaB \uMs^2(\uvr))\rangle$; its small fluctuations only contribute to higher orders~\cite{MM2015}.

When the material parameters fluctuations are small ($\uImr, \uIkr \ll 1$), the approximate solution of the micromagnetic problem~\eqref{eq:energy} can be obtained as a perturbation on top of the ground state that forms in the absence of fluctuations ($\uImr=\uIkr=0$). In our case, the ground state always corresponds to the uniform magnetization state, for which the first and the third terms in~\eqref{eq:energy} are zero because they are proportional to the spatial derivatives of the magnetization. Minimization of the remaining second (Zeeman energy) and last (global anisotropy energy) terms defines the orientation of the uniform ground state magnetization. In the case of $\uKZ=0$ the magnetization is strictly parallel to the external field $\uvH$. The global uniaxial anisotropy will rotate the magnetization away from the direction of the field $\uvH$ by the angle $\utheta$ towards the direction of the anisotropy axis $\uvs$, which can be parametrized via spherical angels {$\utheta_s$ and $\phi_s$}. The corresponding equation for the angle $\theta$ between the vectors $\uvM$ and $\uvH$ follows from~\eqref{eq:Brown}:
\begin{equation}
        \label{eq:StonerWohlfarth}
	\uh\sin\theta - \uQ\sin(\theta_s-\theta)\cos(\theta_s-\theta) = 0,
\end{equation}
where the dimensionless magnetic field equals $\uh=H/(\ugammaB\uMZ)$. It is identical to the one studied by~\citeasnoun{SW1948} and in the general case may exhibit quite a nontrivial hysteresis. In the case when the anisotropy axis is oriented perpendicularly to the external magnetic field ($\theta_s = \pi/2$) the lowest energy solution of~\eqref{eq:StonerWohlfarth} can be written as:
\begin{equation} \label{eq:theta}
	\theta = \arccos\left(\frac{\uh}{\uQ}\right) = \arccos\left(\frac{\uH}{\uHK}\right),
\end{equation}
where $\uHK=2 \uKZ/ (\umuZ \uMZ)$ is the critical field at which saturation (of the coherent rotation of magnetization) occurs and the average magnetization becomes strictly parallel to the applied field $\uvH$. Another interesting case is when the anisotropy axis is directed along $\uvH$ ($\theta_s = 0$). In this case, a square hysteresis loop with a width of $\uHK$ will be observed. Uniform rotation of magnetization by itself is irrelevant to the SANS experiment as it produces no scattering of neutrons, but its interplay with random material parameter fluctuations, as we will see later, makes a substantial impact on the magnetic SANS cross-sections at $\uq\neq0$.

Because the ground state of the considered magnet is always uniform, we introduce another (primed) coordinate system, in which the $\ueZp$ axis is oriented along the ground state magnetization vector. Specifically, the original and the primed coordinate systems are connected by a rotation in the plane containing the magnetic field vector $\uvH$ and the director of the global anisotropy $\uvs$. Denoting the corresponding rotation angle as $\utheta$, this rotation can be described by the rotation matrix $\uRotation(\utheta)$, which relates the coordinates of vectors in the original and primed coordinate systems, \textit{e.g.}\ $\uvsp=\uRotation\uvs$. By construction, among the magnetization vector components in the rotated coordinate system $\uvMp(\uvr) = \{ \uMxp, \uMyp, \uMzp\}$ the first two $\uMxp,\uMyp \ll \uMzp$ are of the first order in $\uImr$ and $\uIkr$ and the last one can be found from the constraint $|\uvM|=\uMs$ as $\uMzp \approx \uMZ + \uMZ \uImr$ up to the terms of the first order in $\uImr$.

Because the SANS cross-sections~\eqref{eq:perpcrossect} and \eqref{eq:parcrossect} are expressed in terms of the Fourier components of the magnetization $\uvFM$, it is convenient to solve Brown's equations directly in Fourier space. This also makes the solution of Maxwell equations for the demagnetizing field easier, but turns products of real-space functions in the effective field and in the Brown's equations into convolutions. They are denoted here as
\begin{equation}
	P \uCONV L = \sum_{\uvqp} P(\uvqp)L(\uvq-\uvqp).
\end{equation}
The algebra of convolutions is commutative, distributive, and associative with respect to multiplication by a constant. It also has an identity element $\delta$ such that $\delta \uCONV P=P$.

Using this notation we can express the Fourier image of the effective field~\eqref{eq:HeffDEF} including all the terms from the energy~\eqref{eq:energy} as:
\begin{eqnarray}
        \uvFHeffp & = &  - \ugammaB \uLZ^2 \uq^2 \uvFMp + \uvHp \delta - \ugammaB \frac{\uvqp[\uvqp \cdot \uvFMp]}{\uq^2} +\nonumber\\
        & &+\ugammaB \widetilde{I_K} \uCONV (\uvdp\uCONV\uvFMp)\uCONV\uvdp  + \ugammaB \uQ (\delta-2\widetilde{\uImr}) \uCONV [\uvFMp \cdot \uvsp] \uvsp,\label{eq:FHeff}
\end{eqnarray}
where the convolution of two vectors $(\uvdp\uCONV\uvFMp)$ is understood as their scalar product with multiplications replaced by convolutions. We also assume here, like in~\citeasnoun{MM2015}, that $\uvFM$ tends to $0$ fast enough as $\uq\rightarrow0$ so that the third term in~\eqref{eq:FHeff} tends to $0$ in this limit. The expression~\eqref{eq:HeffA} is only correct up to the first order in $\uImr, \uIkr \ll 1$. In this order it is, basically, the same as the Fourier image of the effective field in~\citeasnoun{MM2015} apart from the addition of the last term and the fact that it is expressed in the primed coordinate system, whose $\ueZp$ axis may rotate away from the direction of the external field $\uvH$.

Of the three Brown’s equations for each vectorial component of~\eqref{eq:Brown} only two are independent. They also contain convolutions in Fourier space
\begin{eqnarray}
	\uvFHeffZp\uCONV \uFMyp & = & \uvFHeffYp \uCONV \uFMzp , \\
	\uvFHeffZp\uCONV \uFMxp & = & \uvFHeffXp \uCONV \uFMzp.
\end{eqnarray}
Substituting the components of the effective field and the Taylor expansion of the magnetization $\uvMp=\{\uMxpI,\uMypI,\uMZ+\uMZ \uImr\}$ (where the quantities $\uMxpI$, $\uMypI$  are of the same order as $\uImr$, $\uIkr$), Brown's equations become Taylor series themselves.

In zero order we recover the equation~\eqref{eq:StonerWohlfarth}, which was already analyzed. Collecting the first-order terms leads to the following system of equations for the components of the normalized magnetization vector $\uvmp=\uvMp/\uMZ$:
\begin{eqnarray}
        (\uhqp - \uQ \uspysq + \uyqpsq)\uFmypI + (\uxqp \uyqp - \uQ \uspx \uspy)\uFmxpI & = & \uAyp - \uFImr \uyqp\uzqp, \\
	(\uhqp - \uQ \uspxsq + \uxqpsq)\uFmxpI + (\uxqp \uyqp - \uQ \uspx \uspy)\uFmypI & = & \uAxp - \uFImr \uxqp\uzqp,
\end{eqnarray}
where $\uhqp=\uhzp+\uLZ^2 \uq^2 + Q \uspzsq$, $\uAxp=\uFdxp\uCONV\uFdzp\uCONV\uFIkr +\uhxp \uFImr$, $\uAyp=\uFdyp\uCONV\uFdzp\uCONV\uFIkr +\uhyp \uFImr$, $\uvhp = \uvHp/(\ugammaB \uMZ)$, and the components of the $\uvq$~director are $\{\uxqp, \uyqp, \uzqp\}=\uvqp/\uq$. This linear system of equations can be easily solved:
\begin{eqnarray}
    \label{eq:mx1}
    \uFmxpI\!\!\!\! & = &\!\!\! \frac{\uAyp(\uQ\uspx\uspy-\uxqp \uyqp)+\uAxp(\uhqp-\uQ \uspysq+\uyqpsq)-\uFImr \uzqp(\uhqp \uxqp+\uQ\uspy W)}{\uhqpsq-\uQ W^2 +\uhqp(\uxqpsq+\uyqpsq-\uQ(\uspxsq + \uspysq))}, \\ 
    \label{eq:my1}
    \uFmypI\!\!\!\! & = &\!\!\! \frac{\uAxp(\uQ\uspx\uspy-\uxqp \uyqp)+\uAyp(\uhqp-\uQ \uspxsq+\uxqpsq)-\uFImr \uzqp(\uhqp \uyqp-\uQ\uspx W)}{\uhqpsq-\uQ W^2 +\uhqp(\uxqpsq+\uyqpsq-\uQ(\uspxsq + \uspysq))},
\end{eqnarray}
where $W=\uspx\uyqp - \uspy \uxqp$. These two expressions are only different with respect to exchanging $\uX\leftrightarrow\uY$ everywhere (note that $W$ changes sign under such a transformation).

In the absence of the global anisotropy ($\uQ=0$) the primed coordinate system is identical to the original one. Then $\uhxp=\uhyp=0$, $\uhzp=\uh$, $\uhqp|_{\uQ=0}=\uhq=\uh+\uLZ^2\uq^2$ and the resulting expressions for $\uFmxpI$ and $\uFmypI$ coincide with the ones obtained in~\citeasnoun{MM2015}, which in turn have several limiting cases in the earlier micromagnetic SANS theory and in the approach-to-saturation theory.

\section{MAGNETIC SANS CROSS-SECTIONS IN THE LOW-FIELD LIMIT}
\label{sec:lowfield}
In this Section, we analyse the perpendicular scattering cross-sections in the regime of low magnetic fields ($\uH<\uHK$). For brevity and for practical relevance (as illustrated in Section~\ref{sec:expdata}) only the particular case when the anisotropy is directed along the $\ueY$~axis is considered.

When the anisotropy is directed along $\ueY$ the equilibrium magnetization deviates from the field direction by rotating in the $\uY\uO\uZ$ plane around the $\ueX$~axis.
This means that $\uhxp=0$ and $\uspx=0$. Assuming that the anisotropy and saturation magnetization inhomogeneity functions are related via a scalar factor $\kappa$, \textit{i.e.}\ $\uIkr=\kappa \uImr$ and setting $\uImr=\uIr$, noting that in the perpendicular SANS geometry $\uxqp = \uxq = 0$, substituting~\eqref{eq:mx1} and~\eqref{eq:my1} into~\eqref{eq:perpcrossect} and performing an averaging procedure over the defect realizations and the representative volume orientations described in~\citeasnoun{MM2015}, one can obtain the following expression for the macroscopic magnetic spin-misalignment SANS cross-section:
\begin{eqnarray}
    \uSigmadSMPerpp & = & \uV \ubH^2 \uMZ^2 \langle \uFIr^2\rangle \frac{1}{15} \left[\frac{\kappa^2}{\uhqpsq} + \frac{\uzqpsq \kappa^2}{(\uhqp -\uQ \uspysq +\uyqpsq)^2}-\right.\nonumber\\
    & & \left. -\frac{30 \uyqp \uzqp (\uhyp - \uyqp \uzqp)}{\uhqp-\uQ \uspysq +\uyqpsq} + \frac{15 \uzqpsq(\uhyp -\uyqp \uzqp)^2}{(\uhqp-\uQ \uspysq+\uyqpsq)^2} \right]\label{eq:SigmaPY}
\end{eqnarray} 
where $\langle\uFIr^2\rangle$ is the average of the squared Fourier image of the inhomogeneity function $\uFIr$, which for the Gaussian randomly placed inhomogeneities were calculated in~\citeasnoun{MM2015}. We emphasize that the presence of the anisotropy in the $\ueY$~direction rotates the SANS cross-section as a whole and~\eqref{eq:SigmaPY} gives the scattering cross-section in the rotated coordinate system as a function of $\uvqp=\uq\{0,\uyqp,\uzqp\}$.
In the perpendicular scattering geometry [see Fig.~\ref{fig:SANSscheme}(\textit{a})] at low fields ($\uh<\uQ$), this rotation by the angle~\eqref{eq:theta} in the detector plane  can be described using the following rotation matrix:
\begin{equation}
\hat{R} =\left(
    \begin{array}{ccc}
        1 & 0 & 0 \\
        0 & \frac{\uh}{\uQ} & \sqrt{\frac{\uQ^2-\uh^2}{\uQ^2}} \\
        0 & -\sqrt{\frac{\uQ^2-\uh^2}{\uQ^2}} & \frac{\uh}{\uQ} \\
    \end{array}
\right)
\end{equation}
and
\begin{equation} 
\label{eq:stresshcomp}
    \uvs' = \left\{0,\frac{\uh}{\uQ},-\sqrt{\frac{\uQ^2-\uh^2}{\uQ^2}}\right\}, \qquad
    \bm{h}' = \left\{0,\uh \sqrt{\frac{\uQ^2-\uh^2}{\uQ^2}},\frac{\uh^2}{\uQ}\right\}.
\end{equation}
In particular it means that in the absence of an applied field ($\uh=0$), the $\uvsp$ vector is oriented along the $\ueZp$~axis of the primed coordinate system and the average magnetization is directed along the easy axis. By contrast, in a saturating field ($\uh=\uQ$), the field $\uvhp$ is along $\ueZp$, implying that the magnetization is parallel to the applied field. One can obtain $\uSigmadSMPerpInline(\uqy,\uqz)$ (note the absence of primes) from $\uSigmadSMPerppInline(\uqyp,\uqzp)$ by substituting
\begin{equation}
\label{eq:qcomp}
	\uvqp = \uq \left\{0,\uzq \sqrt{\frac{\uQ^2-\uh^2}{\uQ^2}}+\uyq \frac{\uh}{\uQ},\uzq \frac{\uh}{\uQ}-\uyq \sqrt{\frac{\uQ^2-\uh^2}{\uQ^2}}\right\}.
\end{equation}
Formally, the expression~\eqref{eq:SigmaPY}, prior to substitution of the rotation matrix, is valid both in the high- and low-field limits with an appropriate selection of the primed coordinate system. In the high field limit, it can be further simplified as described in the next section.

\section{MAGNETIC SANS CROSS-SECTIONS IN THE HIGH-FIELD LIMIT}\label{sec:highfield}

In the saturation regime ($\uH>\uHK$), the  macroscopic mean magnetization is directed along the external magnetic field. It means that there is no need to rotate the coordinate system and the expressions for the cross-sections become simpler.

\subsection{Perpendicular SANS geometry}\label{subsec:highperp}

\subsubsection{$\uvs \parallel \ueY$}:

Putting $\uhyp = 0$ and $\uspy = 1$ in~\eqref{eq:SigmaPY}, and removing all the primes one obtains the following expression:
\begin{equation} \label{eq:SigmaPYHigh}
	\uSigmadSMPerpY = \frac{\uV \ubH^2 \uMZ^2 \langle\uFIr^2\rangle}{15} \left[ \frac{\kappa^2}{\uhq^2}+ \frac{\uzqsq \kappa^2}{(\uhq-\uQ+\uyqsq)^2} + \frac{15 \uyqsq \uzqsq (2 (\uhq -\uQ + \uyqsq)+\uzqsq)}{(\uhq-\uQ+\uyqsq)^2} \right] .
\end{equation}
In the limiting case of $\uQ = 0$ and after introduction of the polar angle $\alpha$ for the scattering vector in the detector plane, $\uvqPerp=\uq\{0,\sin\alpha,\cos\alpha\}$, it coincides with equation~(49) in~\citeasnoun{MM2015}.

It is convenient~\cite{michels2013} to decompose the spin-misalignment SANS cross-section into the contributions related to anisotropy and saturation magnetization fluctuations:
\begin{equation}
   \label{eq:respFunDef}
   \uSigmadPerp=\uSigmadRES+\uSigmadSMPerp =  \uSigmadRES(\uvqPerp)+\uSH(\uq)\uRHPerp(\uvqPerp,\uh) + \uSM(\uq)\uRMPerp(\uvqPerp,\uh),
\end{equation}
where $\uSigmadRESInline$ is the magnetic-field-independent (nuclear and magnetic) residual SANS cross-section (measured at complete saturation). The function $\uSH(\uq) = \uV \ubH^2 \uMZ^2 \langle\uFIr^2\rangle 2\kappa^2/15$ is called the anisotropy-field scattering function, whereas
$\uSM(\uq)=\uV \ubH^2 \uMZ^2 \langle\uFIr^2\rangle$ is the scattering function of the longitudinal magnetization. Because of the averaging over the scattering volume orientation these functions depend only on the magnitude of the scattering vector. The micromagnetic response functions $\uRHYPerp$ and $\uRMYPerp$, whose superscript here marks the anisotropy axis orientation, have the following form:
\begin{eqnarray}
	\uRHYPerp & = & \frac{1}{2}\left( \frac{1}{\uhq^2}+\frac{\cos^2\alpha}{(\uhq-\uQ+\sin^2\alpha)^2}\right), \\
	\uRMYPerp  & = & \frac{2 \sin^2\alpha \cos^2 \alpha}{(\uhq-\uQ+\sin^2\alpha)}  +\frac{\sin^2\alpha\cos^4 \alpha}{(\uhq-\uQ+\sin^2\alpha)^2}.
\end{eqnarray}
There is no new angular dependence here, but for $\uQ\neq0$ the relative strengths of the isotropic halo and the scattering-angle-dependent terms are modified.

\subsubsection{$\uvs \parallel \ueX$}:
			
A similar [to~\eqref{eq:SigmaPYHigh}] decomposition can also be introduced in this case with the following response functions:
\begin{eqnarray}
    \label{eq:respPerpOX}
    \uRHXPerp & = & \frac{1}{2} \left(\frac{1}{(\uhq-\uQ)^2}+\frac{\cos^2\alpha }{(\uhq+\sin^2\alpha)^2}\right), \\
    \uRMXPerp & = & \frac{2\sin^2\alpha \cos^2\alpha}{(\uhq+\sin^2\alpha)} + \frac{\sin^2\alpha \cos^4\alpha}{(\uhq+\sin^2\alpha)^2}.
\end{eqnarray}
The global anisotropy does not affect $\uRMXPerp$, but only $\uRHXPerp$.

\subsubsection{$\uvs \parallel \ueZ$}:

When the anisotropy axis is oriented along the direction of the magnetic field, the magnetic SANS cross-section is given by:
\begin{equation}
	\uSigmadSMPerpZ = \frac{\uV \ubH^2 \uMZ^2 \langle\uFIr^2\rangle}{15} \left[ \frac{\kappa^2}{\uhqpsq}+ \frac{\uzqsq \kappa^2}{(\uhqp+\uyqsq)^2} + \frac{15 \uyqsq \uzqsq (2 (\uhqp + \uyqsq)+\uzqsq)}{(\uhqp+\uyqsq)^2} \right] .
\end{equation}
Its only difference from the globally isotropic case~\cite{MM2015} is the replacement of $\uhq$ by $\uhqp$, which means that the anisotropy only modifies (increases or decreases, depending on the sign of $\uQ$) the effective external field strength.

\subsection{Parallel SANS geometry}\label{sec:highpar}

Usually the parallel magnetic SANS cross-section is isotropic in the detector plane. However, the presence of a global anisotropy breaks this property. The cross-section can be computed by substituting~\eqref{eq:mx1} and~\eqref{eq:my1} into~\eqref{eq:parcrossect} and averaging over the defect realizations and the representative volume orientations. This yields:
\begin{equation}
	\uSigmadSMPar = \uSH(\uq)\uRHXYPar (\uvqPar,\uh),
\end{equation}
where the response functions $\uRHXYPar$ are no longer isotropic and depend on the polar angle $\beta$ of the scattering vector $\uvqPar = \uq\{ \cos\beta, \sin\beta, 0\}$. They are given by the following expressions:
\begin{equation}
	\uRHXPar = \frac{\uQ \sin^2\beta (2 \uhq-\uQ+2)+(\uhq-\uQ+1)^2}{2 \left(\uhq^2-\uhq \uQ+\uhq-\uQ \sin^2\beta\right)^2},
\end{equation}	
\begin{equation}
	\uRHYPar = \frac{\uQ \sin^2\beta (-2 \uhq+\uQ-2)+(\uhq+1)^2}{2 \left((\uhq+1) (\uhq-\uQ)+\uQ \sin^2\beta\right)^2},
\end{equation}
where the first equation refers to the case $\uvs \parallel \ueX$ and the second expression is for the $\uvs \parallel \ueY$ case. Generally, the SANS cross-section is getting compressed ($\uQ>0$) or expanded ($\uQ<0$) along the anisotropy axis.

When the anisotropy axis coincides with the direction of the external magnetic field ($\uvs\parallel\ueZ$), the parallel spin-misalignment SANS cross-section remains fully isotropic in the detector plane, but the denominator now contains $\uhqp$ instead of $\uhq$ so that $\uSigmadSMParInline=\uSH(\uq)/(2 \uhqpsq)$.

\section{ANALYSIS OF EXPERIMENTAL DATA}\label{sec:expdata}

\subsection{Field-annealed Vitroperm}

\begin{figure}
    \begin{center}
     \includegraphics[width=\textwidth]{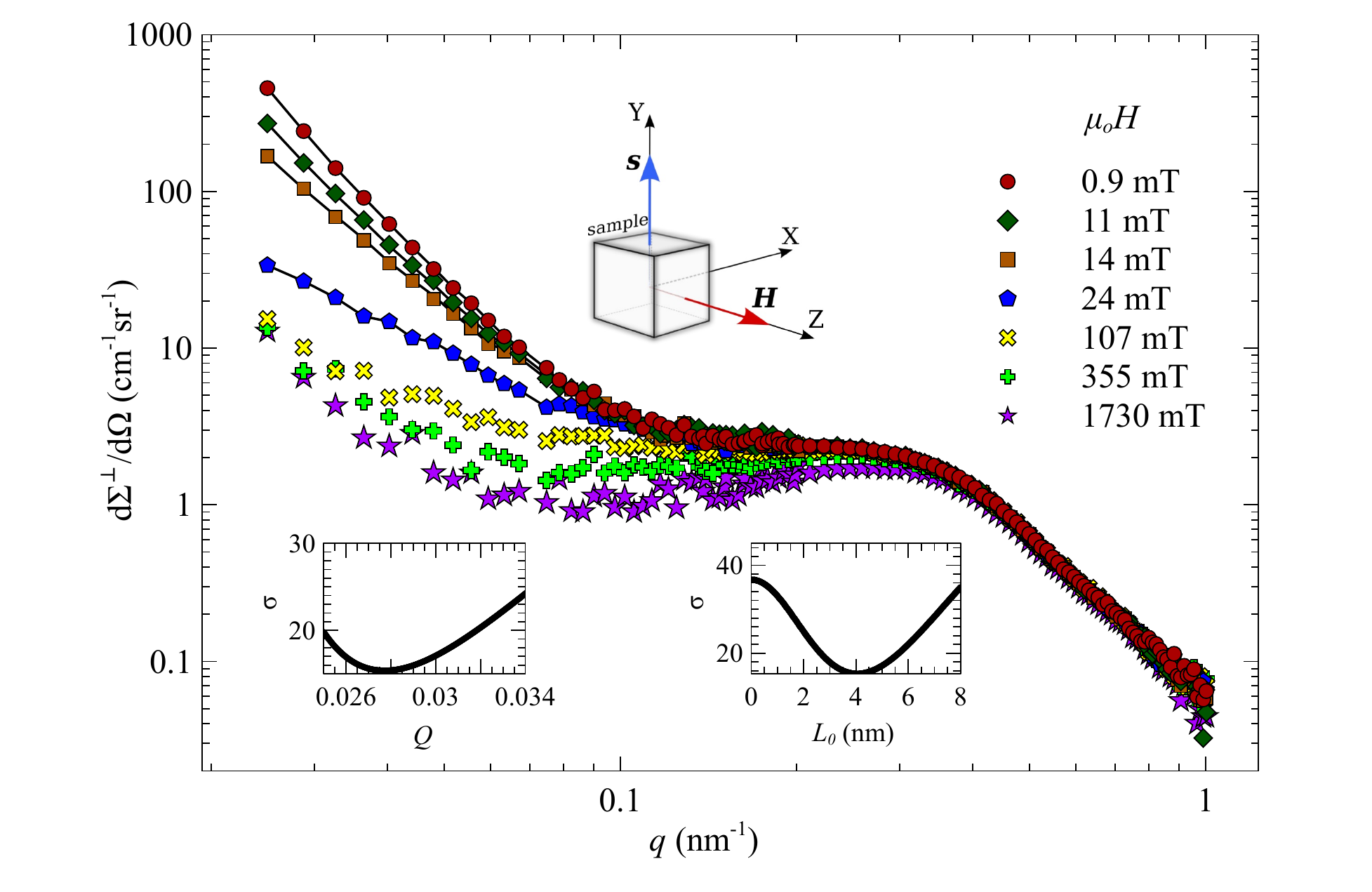}
    \end{center}
    \label{fig:vitroperm}
    \caption{Azimuthally-averaged $\uSigmadPerpInline$ of Vitroperm at selected applied magnetic fields (log-log scale), the experimental data are taken from \citeasnoun{MVW2003}. Solid lines show the fit by~\eqref{eq:respFunDef}, which is also valid in the low-field limit, using the response functions~\eqref{eq:respFunPerpSOYavg} and \eqref{eq:respFunPerpSOYavg1}. The insets show the total mean-square deviation $\sigma$~error for the $\uQ$ and the $\uLZ$ fits described in the text, as well as the relative orientation of the anisotropy axis and the applied magnetic field.}
\end{figure}

One way to induce a global anisotropy into a magnetic nanocrystalline material is to anneal it in a magnetic field. \citeasnoun{MVW2003} and~\citeasnoun{GSHMVW2004} report magnetic SANS data of a field-annealed Vitroperm alloy, which is a two-phase iron-based nanocrystalline soft magnetic material. The azimuthally-averaged perpendicular spin-misalignment cross-section data are shown in Fig.~\ref{fig:vitroperm} as a function of the magnitude of the scattering vector $\uq$ for a number of different values of the applied magnetic field $\uH$. The direction of the anisotropy axis is shown in the insets of Fig.~\ref{fig:vitroperm}. From the Vitroperm hysteresis loop, shown in Fig.~1 of~\citeasnoun{GSHMVW2004}, we can determine that magnetic saturation by coherent rotation occurs at about $\SI{40}{\milli\tesla}$. This means that among the available data the four fields of \num{0.9}, \num{11}, \num{14}, and \SI{24}{\milli\tesla} correspond to the low-field regime, considered in Section~\ref{sec:lowfield}. The data of the remaining three values of the applied field are not sufficient to reliably determine the scattering functions in the high-field regime by fitting~\eqref{eq:respFunDef} with three unknowns. 
In terms of the response functions~\eqref{eq:respFunDef} the spin-misalignment SANS cross-section in the low-field regime in the primed coordinate system~\eqref{eq:SigmaPY} can be represented by:
\begin{eqnarray}
\uRHYPerp &=& \frac{1}{2} \left(\frac{1}{\uhqpsq}+\frac{\uzqpsq}{(\uhqpsq-\uQ\uspysq+\uyqpsq)^2}\right)\\
\uRMYPerp &=& \frac{\uzqpsq (\uyqp\uzqp-\uhyp)^2}{(\uhqp -\uQ\uspysq+\uyqpsq)^2}+\frac{2\uyqp\uzqp(\uyqp\uzqp-\uhyp)}{\uhqp-\uQ\uspysq+\uyqpsq}.
\end{eqnarray}
To interpret the azimuthally-averaged cross-section data, the response functions need to be averaged over $\alpha$ in $\uvqPerp$ as well by computing $\langle\ldots\rangle_\alpha=1/(2\pi)\int_{0}^{2 \pi}(\ldots)\ud\alpha$. The rotation of the cross-section as a whole in the plane of the detector is insignificant for the azimuthal averaging, so that we can integrate directly in the primed coordinate system by setting $\uyqp = \sin\alpha$ and $\uzqp =\cos\alpha$. Substituting the values of $\uvhp$ and $\uvsp$ from~\eqref{eq:stresshcomp} and performing the integration we get:
\begin{eqnarray}
\label{eq:respFunPerpSOYavg}
\langle\uRHYPerp\rangle_\alpha & = & \frac{1}{4}\left(\frac{\uQ^2}{g \sqrt{g(g+Q)}} + \frac{2}{(Q+\lambda^2)^2} \right), \\
\langle\uRMYPerp\rangle_\alpha & = &
\frac{1}{2}\left(\uQ\frac{g(1+\uQ+2\lambda^2)-\uQ\lambda^2(\uQ +\lambda^2)}{g\sqrt{g(g+\uQ)}}-1\right),
\label{eq:respFunPerpSOYavg1}
\end{eqnarray}
where $\lambda=\uLZ\uq$ and $g=-\uh^2+\uQ(\uQ+\lambda^2)$. Taking the known value of the exchange stiffness $\uA=\uC/2=\SI{1e-11}{\joule\per\meter}$ (or $\uLZ = \SI{4.2}{\nano\meter}$) and the saturation magnetization $\umuZ\uMZ=\SI{1.2}{\tesla}$ of Vitroperm only the parameter $\uQ$ in the above functions is unknown. To determine $\uQ$ we have performed a linear regression of the cross-section data for the four chosen values of the field, corresponding to the low-field regime. At each $\uq$, the data were fitted by the expression~\eqref{eq:respFunDef}, obtaining $\uSigmadRESInline(\uq)$, $\uSH(\uq)$, $\uSM(\uq)$, and then the total least-square error of this fit was numerically minimized to obtain the value of $\uQ$. This procedure yields $\uQ=\uQVP=\num{0.0277+-0.0002}$ and the fitted curves are displayed by the solid lines in Fig.~\ref{fig:vitroperm}. Note that the field dependence of the fitted curves is solely determined by the explicit field dependence of the average response functions $\langle\uRHYPerp\rangle_\alpha$ and $\langle\uRMYPerp\rangle_\alpha$. The linear regression parameters $\uSigmadRESInline$, $\uSH$, and $\uSM$ are field-independent. The inset in Fig.~\ref{fig:vitroperm} shows the total error of the fit around the optimum value of $\uQ=\uQVP$.

To verify the self-consistency of the fit, we have repeated the procedure fixing the value of $\uQ=\uQVP$ and treating $\uLZ$ as an adjustable parameter. The minimum of the total least-squares error (shown as an inset in Fig.~\ref{fig:vitroperm}) corresponds to $\uLZ = \SI{4.039+-0.002}{\nano\meter}$. The errors were computed using a Monte-Carlo procedure by adding a random $\pm\SI{5}{\per\centi\meter \per\steradian}$ contribution to the measured $\uSigmadInline$~values and computing the standard deviation of the resulting $\uQ$ and $\uLZ$ across many realizations of this random process.

Using the obtained value of $\uQ$, we can compute the value of the uniaxial anisotropy as $\uKZ = \umuZ \uMZ^2 Q/2 \cong \SI{15900}{\joule\per\meter\cubed}$, which is much larger than $\SI{10}{\joule\per\meter\cubed}$ estimated in~\citeasnoun{GSHMVW2004} based on the analysis of the domain-wall width. However, when the external field is perpendicular to the easy axis, the saturation field is of the order of the anisotropy field. Using the obtained value for $\uKZ$ yields $\umuZ H_K = \frac{2 \uKZ}{\uMZ} = \SI{33.2}{\milli\tesla}$ for the anisotropy field, which is in good agreement with the saturation magnetic field, measured in~\citeasnoun{MVW2003}. Such large uniaxial anisotropy values were previously reported in~\citeasnoun{Herzer_2011} for samples annealed under tensile stress.

\subsection{High-pressure-torsion Nickel}

\begin{figure}
    \begin{center}
     \includegraphics[width=\textwidth]{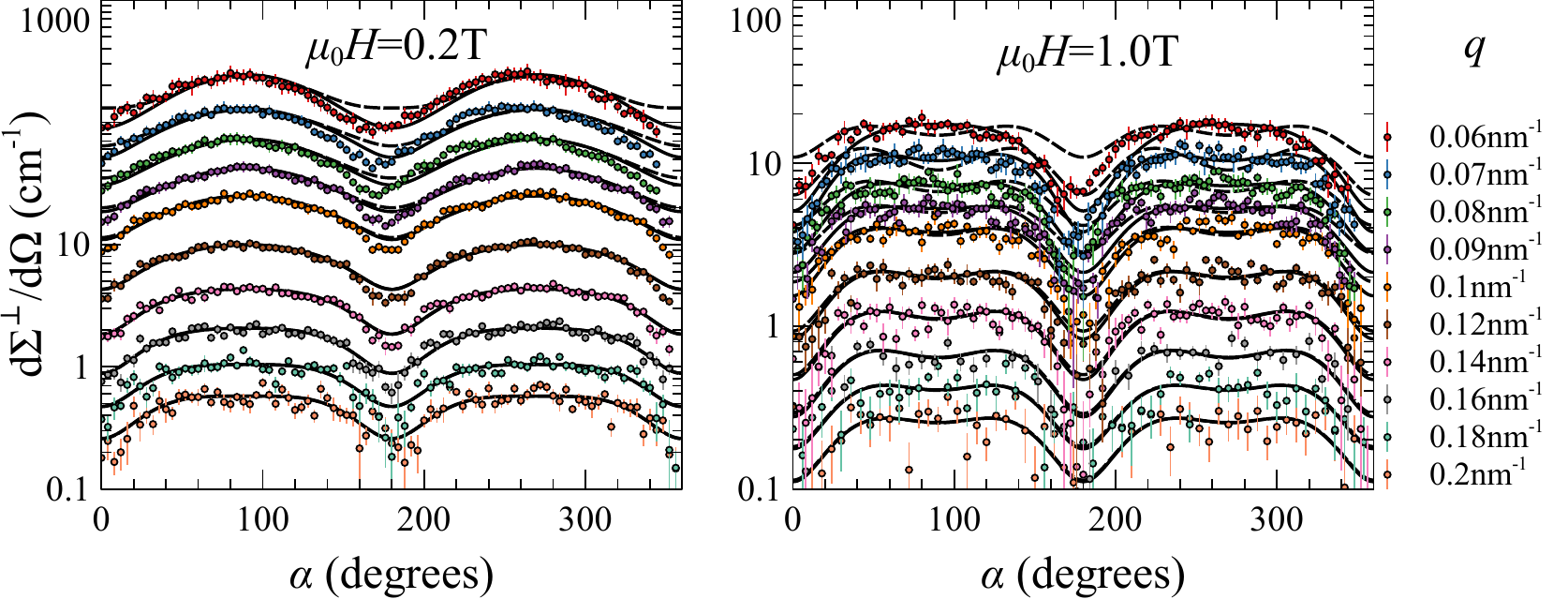}
    \end{center}
    \label{fig:Ni}
    \caption{Angular dependence of the spin-misalignment SANS cross-sections of high-pressure-torsion Ni for two different external fields (see insets). Labeled sets of points, taken from \citeasnoun{OBTATGSMM2021}, correspond to different values of the scattering vector $\uq$. The solid lines show the fit (at their respective value of $\uq$) with optimum value of $\uQ=\uQNi$ and the dashed lines are for $\uQ=0$.}
\end{figure}
Due to its inherent axial symmetry, high-pressure-torsion (HPT) is another way to induce an uniaxial anisotropy in an originally isotropic sample. \citeasnoun{OBTATGSMM2021} report the magnetic SANS cross-section data (shown in Fig.~\ref{fig:Ni}) on a Ni sample, subjected to such a treatment.

In the HPT-Ni experiment the natural anisotropy axis of the sample (the axis of torsion) was aligned with the neutron beam during the SANS experiment and an isotropic parallel SANS cross-section was observed. It is well known that this isotropy indicates that the sample itself is macroscopically isotropic. Yet, as we can see from Section~\ref{subsec:highperp}, despite the fact that there are no new angular terms when the neutron beam is parallel to the anisotropy axis, the cross-section is still slightly modified. These changes can be extracted from the experimental data together with the corresponding anisotropy constant (quality factor $\uQ$) value.

We have fitted the total \textit{perpendicular} SANS cross-section of HPT-Ni (at all $\uq$ and $\uH$~values besides $\umuZ\uH=\SI{0.1}{\tesla}$ as will be explained later) from~\citeasnoun{OBTATGSMM2021} using the following expression:
\begin{equation}
   \label{eq:respFunDefNi}
   \uSigmadPerp=\uSigmadRES(\uvqPerp)+\uSH(\uq)\uRHXPerp(\uvqPerp,\uh) + \uSM(\uq)\uRMXPerp(\uvqPerp,\uh) + S_\mathrm{sin}(\uq,\uH)\sin^2\alpha
\end{equation}
with the response functions from~\eqref{eq:respPerpOX}. The $\sin^2\alpha$~term is introduced and discussed in~\citeasnoun{OBTATGSMM2021} and is beyond the scope of the present work. Also, in~\citeasnoun{OBTATGSMM2021} $\uSH(\uvq)$ and $\uSM(\uvq)$ are assumed to depend on the entire $\uvq$~vector, meaning that each angle $\alpha$ can have its own fitted values of the scattering functions. Here, we assume (as follows from the directional averaging procedure) that these two scattering functions depend only on the magnitude of the scattering vector. Even with such a restriction (and consequently much less fitting freedom) we were able to obtain a comparable quality fit [see the solid lines in Fig.~\ref{fig:Ni} and in Fig.~8 from~\citeasnoun{OBTATGSMM2021}] by minimizing the total error with respect to the $\uQ$~value in the response functions. The fit with $\uQ=0$ (dashed lines in Fig.~\ref{fig:Ni}) is significantly worse, especially at the low values of $\uq$, where the cross-section values are the largest.

The best-fit value for the anisotropy quality factor is $\uQ=\uQNi=\num{0.18}$. For this reason, we had to omit the $\SI{0.1}{\tesla}$ data from the fitting procedure, as (unlike the rest of the data) they do not fall into the high-field regime for which the response functions~\eqref{eq:respPerpOX} were derived. The anisotropy value turns out to be rather large, $\uKZ = \umuZ \uMZ^2 Q/2 \cong 2.6 \times 10^4 \, \mathrm{J/m^3}$ (using $\uMZ = 482 \, \mathrm{kA/m}$). It was silently absorbed into the $\uSH(\uvq)$ and $\uSM(\uvq)$ response functions in~\citeasnoun{OBTATGSMM2021}, but can be revealed using the present more sophisticated theory; and it agrees very well with the value of $1.8 \times 10^4 \, \mathrm{J/m^3}$ estimated in \citeasnoun{bersweiler2021} by a correlation-function analysis on the same specimen.

\section{SUMMARY AND CONCLUSIONS}\label{sec:summary}

The existing micromagnetic SANS theory for spatially inhomogeneous ferromagnets is extended here by including the effect of a nonzero average uniaxial anisotropy with quality factor $Q$. The anisotropy leads to a deviation of the average magnetization from the external magnetic field direction accompanied by the Stoner-Wohlfarth rotational hysteresis. The macroscopically-averaged (over the orientation and realizations of the random material defects in the scattering volume) SANS cross-sections are computed analytically, based on the presented solution of Brown's equations of micromagnetics in the small-misalignment approximation. In view of their simplicity and practical significance, we have compiled the cross-section expressions for several special cases, where the anisotropy axis is either perpendicular or parallel to the magnetic field direction both in the parallel and perpendicular SANS geometry. It follows from Stoner-Wohlfarth theory that for the mutually perpendicular anisotropy axis and the magnetic field direction there is a critical magnetic field at which saturation of the coherent rotation of magnetization occurs. We have analyzed the SANS cross-sections both below and above this rotational magnetic saturation and computed the micromagnetic SANS response functions for the latter case. Some of these expressions exhibit an additional angular dependency on the scattering vector orientation, compared to their previously known $\uQ=0$ limit. The present theory fits well the azimuthally-averaged SANS cross-section of field-annealed Vitroperm alloy and the angular dependence of the spin-misalignment SANS cross-section of high-pressure-torsioned Ni, allowing to determine the respective global anisotropy quality factors. 

\ack{K.L.M.\ acknowledges the support of the Russian Science Foundation under Project No.~RSF 21-11-00325. A.M.\ thanks the National Research Fund of Luxembourg (FNR) for financial support (CORE SANS4NCC project). Y.O. is grateful to Prof. Y. Todaka and Dr. N. Adachi for their help in the experiment and KAKENHI Grant No. 19K05102 for financial support.}

% Tell bibtex which bibliography style to use

\referencelist[vdz_base]

\end{document}